\documentclass[pre,twocolumn,amsmath,tbtags]{revtex4-2}

\usepackage{amssymb}
\usepackage{tikz}
\usepackage{graphicx}
\usepackage{physics}
\usepackage{soul}

\newcommand{\mat}[1]{\mathsf{#1}}

\begin{document}

\title{From a microscopic solution to a continuum description of active particles with a recoil interaction in one dimension}
\author{M. J. Metson}
\author{M. R. Evans}
\author{R. A. Blythe}
\affiliation{SUPA, School of Physics and Astronomy, University of Edinburgh, Edinburgh EH9 3FD, UK}

\begin{abstract}
We consider a model system of persistent random walkers that can jam, pass through each other or jump apart (recoil) on contact. In a continuum limit, where particle motion between stochastic changes in direction becomes deterministic, we find that the stationary inter-particle distribution functions are governed by an inhomogeneous fourth-order differential equation. Our main focus is on determining the boundary conditions that these distribution functions should satisfy. We find that these do not arise naturally from physical considerations, but need to be carefully matched to functional forms that arise from the analysis of an underlying discrete process. The inter-particle distribution functions, or their first derivatives, are generically found to be discontinuous at the boundaries.
\end{abstract}

\maketitle

\section{Introduction}

A central goal in statistical mechanics is to understand how microscopic fluctuations affect the macroscopic behavior of many-body systems. For assemblies of particles in thermal equilibrium with their environment, the procedure is very well established. Working upwards from the microscopic scale, the Boltzmann distribution combined with the principle of detailed balance allows both static and dynamic properties to be predicted  \cite{Reichl2009}. At the mesoscale, one can appeal to free energy minimization and the fluctuation-dissipation theorem to the same ends \cite{Kardar2007}.	

For nonequilibrium systems, the corresponding `bottom-up' and `top-down' approaches are still under development, particularly in the context of active matter formed of internally-driven particles that seek to maintain a persistent motion \cite{Cates2022}. The top-down approach is perhaps more straightforward, and works in the spirit of Landau free-energy theory by appealing to a low-order expansion in the physical fields of interest with the addition of physically-motivated noise to gain insights into the dynamics \cite{Toner1998,Marchetti2013}. However, in this approach one loses connection to the properties of individual particles and in particular how they contribute to the noise. By working from an explicit microscopic model, the bottom-up approach furnishes the information that a top-down approach cannot provide, at the expense of extra difficulty (see e.g.~\cite{Tailleur2008,Thompson2011,Farage2015,Steffenoni2017,OLaighleis2018}). One way to address this difficulty is to start with a Langevin or Fokker-Planck equation that describes a single particle's motion in response to a coarse-grained density field. A self-consistent formulation then arises by integrating over an assembly of particles to obtain the density field \cite{Dean1996}.

In this work, we focus on the construction of equations that appropriately describe the stochastic dynamics of interacting active particles at the microscopic scale, these lying at the heart of any bottom-up approach. Persistent particles, that is, those that attempt to maintain a constant velocity over extended times, serve as a paradigm for out-of-equilibrium matter, with a range of applications from heat transport in turbulent fluids \cite{Taylor1922} and assemblies of self-phoretic particles, motile microorganisms such as \textit{Escherichia coli} and macro-organisms like birds and fish \cite{Viksek2012,Romanczuk2012,Marchetti2013,Elgeti2015,Bechinger2016,Julicher2018}. Specifically, we consider a model system, introduced in \cite{shortshock}, where persistent hard-core particles may jam, pass through each other or recoil on contact and find that there are many subtleties relating to the boundary conditions in the continuum limit. 

The properties of a single persistent particle are by now well understood (see e.g.~\cite{Angelani2014,Malakar2018,Demaerel2018,Hartmann2020,Malakar2020,Mori2020,Mori2021,DeBruyne2021,Singh2021b,Garcia2021}). Here we focus on run-and-tumble motion, which is inspired by that of the \textit{E.~coli} bacterium. It comprises runs in a fixed direction for a time that is drawn from a roughly exponential distribution \cite{Korobkova2004,Saragosti2012}, after which a new direction emerges through the bacterium tumbling (i.e., rotating) for some period of time \cite{Berg2004}. In one dimension, a persistent particle that undergoes velocity reversals as a Poisson process is described by the telegrapher equations \cite{Goldstein1951,Rosenau1993,Masoliver1993}, and the stationary distribution of particle positions can be found by appealing to flux balance conditions \cite{Schnitzer1993}.

Understanding the consequences of interactions between persistent particles at the microscopic scale has been more challenging. Perhaps the simplest interaction is hard-core exclusion, which in passive (equilibrium) systems serves only to reduce the volume available to particles to explore: the statistical weights of the accessible microstates remain unchanged. By contrast, a hard-core repulsion between persistent particles induces an effective attraction \cite{Slowman2016} which can lead to particles clustering \cite{Soto2014,Sepulveda2016,Kourbane2018,Zhang2019,Metson2020}, consistent with the prediction of motility-induced phase separation \cite{Cates2015}.

So far, exact results have been obtained only for two particles with a hard-core exclusion interaction. These include the statics and dynamics of the inter-particle distribution function when tumbling is instantaneous \cite{Slowman2016,Mallmin2019} and also in the presence of additional thermal noise \cite{Das2020}. The stationary inter-particle distribution function is also known when tumbling lasts for a finite time drawn from an exponential distribution \cite{Slowman2017}. The general many-body problem remains challenging. Although it is straightforward to generalize the telegrapher equations to multiple particles, it is not obvious what boundary condition corresponds to a hard-core exclusion constraint. Whilst it is clear that the spatial particle current must vanish at points of contact, due to particles being unable to pass through each other, the probability flux between different velocity configurations is not subject to any such constraints. Thus one needs additional information to fully specify the boundary conditions.

Here we further our understanding by going beyond a hard-core repulsion, which causes persistent particles that are approaching each other to jam against each other, to a contact interaction where particles stochastically recoil from each other. This interaction is inspired by that observed in the species \textit{Pyramimonas octopus}, in which context it has been described as `shocking' \cite{Wan2016,Wan2018}. We avoid this terminology here so as not to confuse with discontinuous density profiles in fluids, and instead refer to a \emph{recoil} interaction. By being able to vary the distribution of recoil lengths, as well as the relative contributions from jamming or particles passing through each other, we can build up a more complete picture of how a persistent particle's behavior is affected by the contact dynamics, something that has been more fully explored in the absence of persistence \cite{Schutz1996,Alimohammadi1998,Jara2007,Bernardin2017}. This takes us beyond previous microscopic models and is a first step towards a fundamental understanding of nontrivial active interactions, such as those found in biological contexts, where it is difficult to model systems in their full complexity.

Our main finding, which appears in Section~\ref{sec:match}, is that the stationary distribution for a pair of interacting persistent particles is inherently singular at zero particle separation, with discontinuities in the probability density or its derivative arising in different velocity configurations. When there is recoil without jamming, as considered in \cite{shortshock}, these discontinuities can be handled within a fairly na\"{\i}ve treatment, whereas in the general case much more care is needed. Our approach begins in Section~\ref{sec:moddef} with a discretized version of the dynamics. This avoids uncertainty in the boundary conditions, as these are fully specified at the microscopic scale in terms of events that can and cannot occur. We then show in Section~\ref{sec:sss} that the stationary inter-particle distribution functions can be solved on the lattice by applying the kernel method \cite{Banderier2002,Prodinger2004,Slowman2016}. Although this approach yields tractable expressions for their generating functions, direct inversion leads to forms that do not lend themselves to an easy interpretation. Instead, we find in Section~\ref{sec:ctm} that we can take the continuum limit at an earlier stage in the calculation. Importantly, this yields a pair of decoupled fourth-order differential equations for the inter-particle distribution functions, Eq.~(\ref{de}) below.
The general solution follows by integration as described in Section~\ref{sec:bulk}.

To deal with the boundaries, it turns out that the most fruitful approach is to return to the original master equations for the lattice process and establish in Section~\ref{sec:bdy} the appropriate functional forms that the stationary solutions must adopt in the boundary regions. The constants of integration that appear in the general solution in the bulk region are then fixed in Section~\ref{sec:match} by matching to the boundary behavior: this leads to the main finding highlighted above. The fact that the decoupled equations in the bulk are fourth order means that one can match independently the limiting value of the distribution and its derivative at each boundary. Moreover, we find that, near the boundaries, the inter-particle distribution function for following particles varies rapidly over a region of size $\frac{1}{\sqrt{L}}$ where $L$ is the lattice size. This feature thus sharpens to a step function in the continuum limit, $L\to\infty$. Curiously, we find that the derivative of the stationary distribution for approaching particles is discontinuous at each boundary, further indicating that establishing appropriate boundary conditions on the multi-particle telegrapher equations is a delicate exercise.  The case where particles cannot recoil, but only jam or pass through each other when they meet, turns out to be a special case, treated in Section~\ref{sec:special}. We find that delta function contributions of the type previously identified in \cite{Slowman2016} appear only when jamming is the dominant interaction. Since the derivation of the inter-particle distribution involves a number of steps, we collect together the main results in Section~\ref{sec:summary}. We then conclude in Section~\ref{sec:disco} by viewing these findings in the light of the general question of identifying boundary conditions on stochastic equations for interacting particles that are driven out of equilibrium.

\section{Model definition}
\label{sec:moddef}

\begin{figure*}
    \begin{center}
        \includegraphics[scale=1.1]{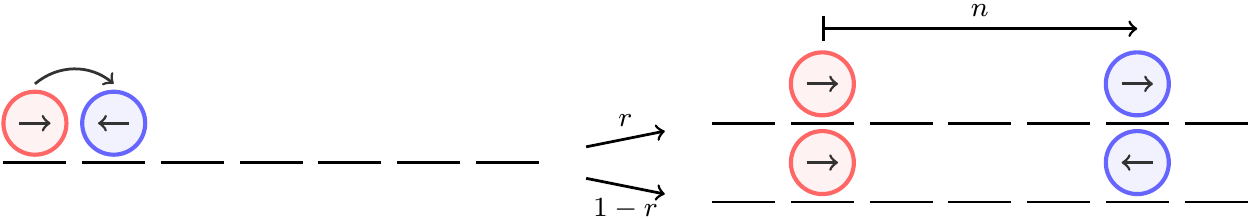}
    \end{center}
    \caption{An illustration of the recoil interaction. The left particle successfully hops whilst the right particle is displaced to a separation of $n$ sites according to $\Phi(n)$. In addition, its velocity is reversed with probability $r$.}
    \label{fig::shock_schematic}
\end{figure*}

The dynamics considered in this work is defined as follows. Two particles inhabit a periodic lattice of $L$ sites in one dimension. Each particle has a direction of motion, denoted $+$ or $-$, and hops to the adjacent site in that direction as a Poisson process with rate 1 (thereby setting the unit of time). Similarly, each particle reorients as a Poisson process with rate $\alpha$. Here, reorientation means choosing one of the two available directions of motion with equal probability and thus the rate at which a particle changes direction is $\omega=\frac{\alpha}{2}$. In addition, the particles recoil upon colliding with each other: if particle 1 hops onto the site occupied by particle 2, the latter is instantaneously displaced parallel to particle 1's direction of motion to a new separation of $n$ sites according to some distribution $\Phi(n)$, where $0 < n < L$. After recoiling, the displaced particle's velocity is reversed with probability $r$; notable cases include $r=0$, corresponding to no reversals, and $r=\frac{1}{2}$, corresponding to velocity randomization. We will refer to spontaneous changes of direction as \emph{reorientation} and those induced by recoiling as \emph{reversal}.

A schematic representation of the recoil dynamics is given in Fig.~\ref{fig::shock_schematic}. This model is a generalization of the persistent exclusion process as described in \cite{Slowman2016}, where the case $\Phi(n)=\delta_{n,1}$ with $r=0$ corresponds to hard-core exclusion. The more general interaction considered here---like the exclusion interaction in \cite{Slowman2016}---breaks detailed balance, thus giving rise to non-equilibrium probability currents \cite{Mallmin2019}.

There are four sectors to consider in this model, corresponding to the directional combinations of the particles: $++$, $+-$, $--$, and $-+$. In what follows, we aim to calculate the stationary probability distributions for the particle separations in each of the four sectors, as well as the net distribution obtained from summing over the four sectors. Due to symmetry considerations, it will suffice to calculate the distributions in the $++$ and $+-$ sectors.

\section{Stationary-state solution}
\label{sec:sss}

Our starting point in the analysis is the set of master equations that describe the evolution of the system. In this section, we write these out and show how they are solved using the kernel method.

\subsection{Master equations}

The state of the system is given in terms of the separation between the two particles, $n$, and each particle's direction of motion, $\sigma_i=\pm$, where $i=1,2$ and $+$ indicates movement to the right, and $-$ movement to the left. For separations $1<n<L-1$, we have for two particles moving to the right that
\begin{multline}
\label{me1}
\dot{P}_{++}(n) = [P_{++}(n-1)-2P_{++}(n)+P_{++}(n+1)] + {}\\
\shoveleft\quad \omega[ P_{-+}(n) + P_{+-}(n) - 2 P_{++}(n) ] + {}\\
\shoveleft\quad r [ P_{+-}(1) \Phi(n) + P_{-+}(L-1) \Phi(L-n) ] + {}\\
\shoveleft\quad (1-r) [ P_{++}(1) \Phi(n) + P_{++}(L-1) \Phi(L-n) ] \;.
\end{multline}
The terms in the first line account for particle hops. Those in the second line arise from reorientations that result in a change of direction, events that occur at rate $\omega = \frac{\alpha}{2}$. The terms in the third line derive from recoil combined with a subsequent reversal; and in the final line to recoil without a reversal. The first line needs to be modified at separations $n=1$ and $n=L-1$, as it is not possible to enter these configurations from $n=0$ or $n=L$. However, it is still possible to exit these configurations by a particle hop, due to the recoil that then takes place. In the discrete setting, we can accommodate the hard-core exclusion constraint by imposing the boundary conditions
\begin{equation}
\label{mebc1}
P_{++}(0) = P_{++}(L) = 0 \;.
\end{equation}

For a pair of approaching particles we obtain, by similar considerations,
\begin{multline}
\label{me2}
\dot{P}_{+-}(n) = 2[P_{+-}(n+1) - P_{+-}(n)] + {}\\
\shoveleft\quad \omega[ P_{++}(n) + P_{--}(n) - 2P_{+-}(n) ] + {} \\
\shoveleft\quad \{ r [P_{++}(1) + P_{--}(1)] + 2(1-r) P_{+-}(1) \} \Phi(n)
\end{multline}
for $n<L-1$. We can extend this equation to $n=L-1$ by imposing the boundary condition
\begin{equation}
\label{mebc2}
P_{+-}(L) = 0 
\end{equation}
to account for the fact that the $n=L-1$ state cannot be entered from $n=L$. There is no corresponding boundary condition for $P_{+-}(0)$, since this term never enters into any of the master equations. The value of $P_{+-}(1)$ determines the overall normalization, and can therefore be set arbitrarily. We will find below that, in the continuum limit, the stationary solutions do not necessarily approach the imposed boundary values smoothly, and that some care is required in handling the behaviour at the boundaries.

To obtain the stationary distributions $P_{\sigma_1\sigma_2}(n)$, we note some important symmetries. First, there is an invariance under particle relabelling,
\begin{equation}
P_{\sigma_1\sigma_2}(n) = P_{\sigma_2\sigma_1}(L-n) \;,
\end{equation}
since the gap between particle 1 and 2 is indistinguishable from the gap between particle 2 and 1. As the dynamics are invariant under a parity transformation, we further have the symmetry
\begin{equation}
P_{\sigma_1\sigma_2}(n) = P_{\bar\sigma_2\bar\sigma_1}(n)
\end{equation}
where $\bar\sigma$ is the direction opposite to $\sigma$. Combining these two symmetries, we find that
\begin{align}
\label{s1}
P_{++}(n) &= P_{++}(L-n)  = P_{--}(n) = P_{--}(L-n) \\
\label{s2}
P_{+-}(n) &= P_{-+}(L-n) \;.
\end{align}
Thus it is sufficient to solve for $P_{++}(n)$ and $P_{+-}(n)$ to find the stationary distribution across all four velocity sectors.

\subsection{Generating functions and kernel method}

As in \cite{Slowman2016,Slowman2017} the master equations can be solved exactly by introducing the generating functions 
\begin{align}
G_{\sigma_1 \sigma_2}(s) &= \frac{1}{P_{+-}(1)} \sum_{n=1}^{L-1} P_{\sigma_1 \sigma_2} (n) s^n \\
\tilde\Phi(s) &= \sum_{n=1}^{L-1} \Phi(n) s^n \;,
\end{align}
and applying the kernel method \cite{Banderier2002,Prodinger2004}. Here we find it convenient to normalize by $P_{+-}(1)$.

The first step is to sum over (\ref{me1}) and (\ref{me2}) and set the left-hand sides equal to zero to ensure stationarity. Exploiting the symmetries (\ref{s1}) and (\ref{s2}), we find
\begin{multline}
	\label{gf1}
    \left(s + s^{-1} - 2(1+\omega)\right) G_{++}(s) + \omega [G_{+-}(s) + G_{-+}(s)] = {} \\ 
     (1+s^L)\kappa_L - [r + (1-r)\kappa_L] [ \tilde\Phi(s) + s^L \tilde\Phi(s^{-1}) ]
\end{multline}
and
\begin{multline}
	\label{gf2}
    \left(s^{-1} - (1+\omega)\right)G_{+-}(s) + \omega  G_{++}(s) = {}\\
    1- [r\kappa_L+(1-r)] \tilde\Phi(s) \;.
\end{multline}
Here, we have introduced the quantity
\begin{equation}
\label{kappa}
\kappa_L = \frac{P_{++}(1)}{P_{+-}(1)}
\end{equation}
that features prominently in the ensuing analysis. Importantly, this ratio depends on the system size $L$, which we have highlighted with the subscript on $\kappa$.

We further require an equation for the generating function $G_{-+}$. This we obtain by appealing to the symmetries (\ref{s1}) and (\ref{s2}), which imply that
\begin{equation}
G_{\sigma_1\sigma_2}(s^{-1}) = s^{-L} G_{\sigma_2\sigma_1}(s) \;.
\end{equation}
Then putting $s\to s^{-1}$ in (\ref{gf2}), and multiplying by $s^L$, we find
\begin{multline}
	\label{gf3}
    (s-(1+\omega))G_{-+}(s) + \omega G_{++}(s) = {} \\
    P_{+-}(1) s^L - [r\kappa_L+(1-r)] s^L \tilde\Phi(s^{-1}) \;.
\end{multline}

Equations~(\ref{gf1}), (\ref{gf2}) and (\ref{gf3}) comprise a linear system for the unknown generating functions $G_{++}(s)$, $G_{+-}(s)$ and $G_{-+}(s)$. Their solution can be written as
\begin{equation}
\label{linsol}
\frac{K(s)}{s^2} \left(\begin{array}{c} G_{++}(s) \\ G_{+-}(s) \\ G_{-+}(s) \end{array}\right) = \mat{A}(s) \vec{b}(s)
\end{equation}
in which the \emph{kernel}
\begin{equation}
K(s) = (1+\omega)(s-z)(s-z^{-1})(s-1)(1-s)
\end{equation}
involves the two reciprocal roots $z$ and $z^{-1}$ of the quadratic equation
\begin{equation}
\label{z}
z^2 - 2(1+\omega) z + 1 = 0 \;.
\end{equation}
On the right-hand side of (\ref{linsol}) we have
\begin{widetext}
\begin{align}
\label{Amat}
\mat{A}(s) &= \left(\begin{array}{ccc}
    \mu(s)\nu(s) & -\omega\mu(s) & -\omega\nu(s) \\
    -\omega\mu(s) & \mu(s)(\mu(s)+\nu(s))-\omega^2 & \omega^2 \\
    -\omega\nu(s) & \omega^2 & \nu(s)(\mu(s)+\nu(s))-\omega^2
\end{array}\right) \\
\label{bvec}
\vec{b}(s) &=\left(\begin{array}{c} 
        (1+s^L)\kappa_L - [r+(1-r)\kappa_L] [ \tilde\Phi(s) + s^L \tilde\Phi(s^{-1})] \\
        P_{+-}(1) - [\kappa_L + (1-r)] \tilde\Phi(s) \\
        P_{+-}(1)s^L - [\kappa_L + (1-r)] s^L \tilde\Phi(s^{-1})
 \end{array}\right)
\end{align}
\end{widetext}
in which
\begin{align}
\mu(s) &= s - (1+\omega) \\
\nu(s) &= s^{-1} - (1+\omega) \;.
\end{align}

The kernel method \cite{Banderier2002,Prodinger2004} furnishes an explicit expression for the ratio $\kappa_L$. The basic idea is to ensure that the left- and right-hand sides of (\ref{linsol}) both vanish in the same way at each of the kernel's roots. For example, $K(s) \sim (s-z)$ as $s\to z$ and $K(s) \sim (s-1)^2$ as $s\to 1$. As we now show, this behavior is reproduced by the right-hand side (\ref{linsol}) only when $\kappa_L$ is suitably chosen.

To this end, we first observe that
\begin{equation}
\mat{A}(1) \vec{b}(1) = \lim_{s\to 1} \frac{\mat{A}(s) \vec{b}(s)}{s-1} = 0 \;.
\end{equation}
This implies that the right-hand side already has the desired $(s-1)^2$ behavior as $s\to1$. In order to reproduce the $(s-z)$ and $(s-z^{-1})$ behavior in the vicinity of the roots $s=z$ and $s=z^{-1}$, we find that we must have
\begin{equation}
\mat{A}(z) \vec{b}(z) = \mat{A}(z^{-1}) \vec{b}(z^{-1}) = 0 \;.
\end{equation}
At both roots, we find the equality is satisfied as long as
\begin{equation}
    \label{eq::prob_ratio}
    \kappa_L = \frac{\omega(z^L-1) + r\mu(z) \tilde\Phi_+(z) + (1-r)\omega \tilde\Phi_-(z)}{\mu(z)(z^L+1) - r\omega \tilde\Phi_-(z) - (1-r)\mu(z) \tilde\Phi_+(z)} \;.
\end{equation}
Here, we have found it convenient to introduce the generating functions of the symmetric and anti-symmetric components of the recoil distribution,
\begin{equation}
\label{Phitpm}
\tilde\Phi_{\pm}(s) = \sum_{n=1}^{L-1} [\Phi(n) \pm \Phi(L-n)] s^n = \Phi(s) \pm z^L \Phi(s^{-1}) \;.
\end{equation}
Note that the expression (\ref{eq::prob_ratio}) is invariant under the replacement $z \to \frac{1}{z}$, since every term in the numerator and denominator is multiplied by $(-z^{-L})$ under this transformation. This means that it does not matter which root of (\ref{z}) we use in the subsequent analysis. We generally choose the smaller of the two roots.

In principle, one can find explicit expressions for the stationary distributions $P_{\sigma_1\sigma_2}(n)$ by multiplying both sides of (\ref{linsol}) by $\frac{s^2}{K(s)}$, and expanding the right-hand side in powers of $s$. The coefficient of $s^n$ then furnishes  $P_{\sigma_1\sigma_2}(n)$, up to the overall normalization $P_{+-}(1)$.  Given the complexity of (\ref{Amat}) and (\ref{bvec}), we anticipate that the resulting expressions are unwieldy and hard to interpret, although they are likely to simplify in the continuum limit.  Thus we focus in the following on the most efficient path to these limiting expressions, which involves taking the limit at an earlier stage in the calculation.

\begin{figure}[tb]
    \begin{center}
        \includegraphics[width=\linewidth]{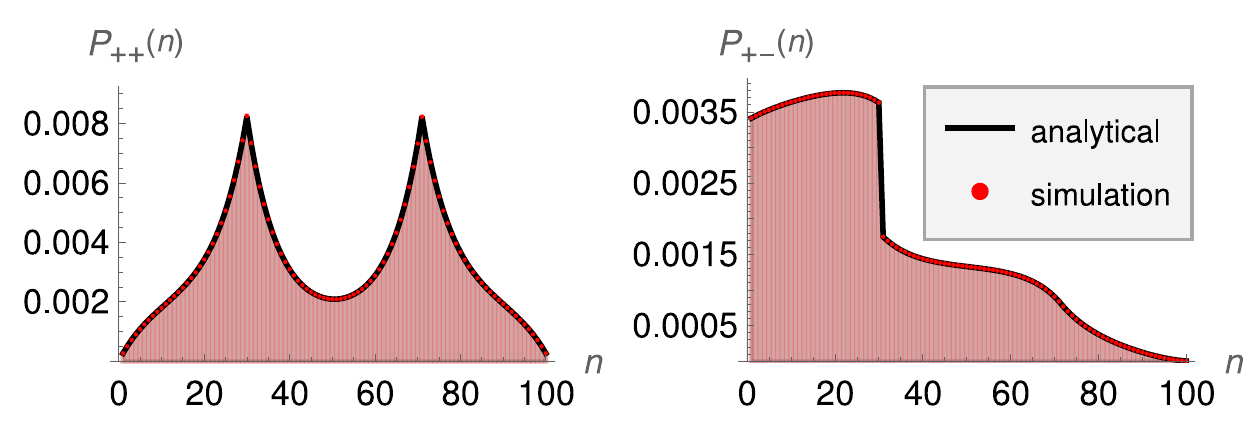}
    \end{center}
    \caption{The discrete distributions $P_{++}(n)$ and $P_{+-}(n)$ from Appendix~\ref{sec:urgh} plotted against simulation data for $\omega=\frac{1}{100}$ on a $101$-site lattice for the simple but nontrivial choice of recoil distribution $\Phi(n)=\delta_{n,30}$ and $r=\frac{1}{2}$. The normalization is such that all four sectors sum to unity. As demonstrated, the solution is indistinguishable from the simulation data.}
    \label{fig::test_distribution_discrete}
\end{figure}

Nevertheless, if one seeks only to evaluate the distributions numerically, one can use the explict formul\ae\ given in Appendix~\ref{sec:urgh} that are obtained through the procedure outlined above. In Fig.~\ref{fig::test_distribution_discrete} we see that these expressions agree perfectly with distributions obtained using direct Monte Carlo simulations of the model dynamics. We note in particular the highly nontrivial forms of the distributions which result from even the simplest choice of recoil dynamics.

\section{Continuum limit}
\label{sec:ctm}

The continuum limit is obtained in the same way as set out in \cite{Slowman2016}. We take both the lattice spacing $a$ and the reorientation rate $\omega$ to zero as 
\begin{equation}
a = \frac{1}{L} \qq{and} \omega = \frac{1}{L\xi} \qq{as $L\to\infty$}
\end{equation}
where we recall $L$ is the system size. Here, $\xi$ is a dimensionless \emph{persistence length}, equal to the fraction of the lattice that is covered by a single particle between two spontaneous changes of direction. This quantity is taken to be fixed in the $L\to\infty$ limit in which space becomes continuous. In this limit, each particle moves ballistically with unit velocity over a distance drawn from an exponential distribution with mean $\xi$ between reorientations, unless its passage is curtailed by encountering the other particle.

We consider a class of recoil distributions $\Phi(n)$ that comprises three parts. The first contribution is a distribution $\rho(x)$ that is normalized on the interval $0\le x \le 1$, and is differentiable at the boundary points. It can, however, be discontinuous or have delta function contributions away from these points. The analysis presented in \cite{shortshock} applies only to distributions with this single contribution.

Here we extend to the case where there are also delta contributions at the boundaries. More precisely,
\begin{equation}
\label{Phiuvw}
\Phi(n) = u \delta_{n,1} + v \delta_{n,L-1} + w \frac{\rho({\textstyle \frac{n}{L}})}{L}
\end{equation}
in which $u$, $v$ and $w$ are probabilities that sum to unity. Defining $x=\frac{n}{L}$, the limiting form of this distribution is
\begin{equation}
    \phi(x) = \lim_{L\rightarrow\infty} L\Phi(Lx) = u\delta(x) + v\delta(1-x) + w\rho(x) .
\end{equation}
With this choice, the dynamics of the pair of particles is as follows. When one particle attempts to hop on top of the other, it jams (i.e., remains at separation $n=1$) with probability $u$, exchanges places with the other particle with probability $v$, and, with probability $w=1-u-v$, causes the other particle to recoil by a distance $x=\frac{n}{L}$ which is sampled from the distribution $\rho(x)$. We recall that the recoiling particle has a probability $r$ of reversing when it reaches its destination. Thus this provides a means of unjamming even when $u=1$ and $v=w=0$. The special case of a pair of hard-core particles that do not recoil on contact, which was solved in \cite{Slowman2016}, is recovered with $u=1$, $r=v=w=0$. 

To obtain the continuum limit of (\ref{linsol}) we first note that we can write
\begin{equation}
\frac{K(s)}{s^2} = (1+\omega)(1-s)(s^{-1}-1) [ 2\omega- (1-s)(s^{-1}-1)] \;.
\end{equation}
For a generating function $\tilde{f}(s) = \sum_n f(n) s^n$ we have the correspondence
\begin{equation}
(1-s)(s^{-1}-1) \tilde{f}(s) \rightleftarrows f(n-1) - 2f(n) + f(n+1) \;.
\end{equation}
We recognize the right-hand side as the action of a second-order finite-difference operator, $\Delta^2$, on the function $f(n)$. This implies that the generating functions on the left-hand side of (\ref{linsol}) can be inverted back to probabilities as
\begin{equation}
\label{discrete4}
\frac{(1+\omega)\Delta^2 ( 2\omega - \Delta^2 )}{P_{+-}(1)} \left(\begin{array}{c} P_{++}(n) \\ P_{+-}(n) \\ P_{-+}(n) \end{array}\right) \;.
\end{equation}
Thus we have transformed the original set of master equations, (\ref{me1}) and (\ref{me2}), which are coupled and involve finite differences up to second order, into a set of decoupled fourth-order equations. This decoupling implies that we can now solve the distribution in each sector separately. Furthermore, since $P_{-+}(n)=P_{+-}(L-n)$, we need only to solve for $P_{++}(n)$ and $P_{+-}(n)$. Introducing now the limiting form of the stationary distribution as
\begin{align}
\label{p}
p(x) &= \lim_{L\to\infty} \frac{P_{++}(Lx)}{P_{+-}(1)}  \\
\label{q}
q(x) &= \lim_{L\to\infty} \frac{P_{+-}(Lx)}{P_{+-}(1)}
\end{align}
we find that, as $L\to\infty$, (\ref{discrete4}) becomes
\begin{equation}
\label{lhs}
\frac{1}{L^3 \xi} \dv[2]{x} \left( 2 - \frac{\xi}{L} \dv[2]{x} \right) \left(\begin{array}{c} p(x) \\ q(x)  \end{array}\right) \;.
\end{equation}

We now turn our attention to the right-hand side of (\ref{linsol}). By a similar argument to that above, we have the correspondences
\begin{align}
\mu(s) \tilde{f}(s) &\rightleftarrows -\frac{1}{L\xi} (1+L\xi{\Delta}_-) f(n) \\
\nu(s) \tilde{f}(s) &\rightleftarrows -\frac{1}{L\xi} (1-L\xi{\Delta}_+) f(n)
\end{align}
in which ${\Delta}_{\pm} f(n) = \pm[ f(n\pm1) - f(n)]$ are first-order finite-difference operators.  For large $L$, and sufficiently far from the boundaries, the first two rows on the right-hand side of (\ref{linsol}) become
\begin{equation}
\label{rhs}
- \frac{1}{L^3\xi} \left(\begin{array}{c} f_p(x) \\ f_q(x)  \end{array}\right)
\end{equation}
in which
\begin{multline}
\label{fp}
f_p(x) = -\xi \left[ r + (1-r) \kappa_L \right] w \dv[2]{\rho_+}{x} + {} \\
\left[ r \kappa_L + (1-r)  \right] w \dv{\rho_-}{x} + \frac{1}{\xi} \left[1 + \kappa_L \right] w \rho_+(x)
\end{multline}
and
\begin{multline}
\label{fq}
f_q(x) = \left[1 + \kappa_L \right]  w \dv{\rho_+}{x} + \left[ r \kappa_L+(1-r)\right] w \dv{\rho_-}{x} + {}  \\ 
\frac{1}{\xi} \left[1 + \kappa_L \right] w \rho_+(x) \;.
\end{multline}
In these equations, $\rho_{\pm}(x)$ are the (anti-)symmetric components of the recoil distribution in the bulk, that is,
\begin{equation}
\rho_{\pm}(x) = \rho(x) \pm \rho(1-x) \;.
\end{equation}
Equating (\ref{lhs}) and (\ref{rhs}), we find that in the bulk, the limiting forms of the probability distributions $p(x)$ and $q(x)$ are governed by the fourth-order equation
\begin{equation}
\label{de}
\dv[2]{x} \left( \frac{\xi}{L} \dv[2]{x} -2 \right) \left(\begin{array}{c} p(x) \\ q(x)  \end{array}\right) = \left(\begin{array}{c} f_p(x) \\ f_q(x)  \end{array}\right) \;.
\end{equation}
It is tempting to drop the fourth derivative since its prefactor is of order $\frac{1}{L}$ relative to that of the second derivative. However, there are situations where this derivative cannot be neglected, in particular, at the boundaries and when there are discontinuities in the recoil distribution in the bulk. 

\section{Solution in the bulk}
\label{sec:bulk}

We now solve (\ref{de}) for an arbitrary combination of the probabilities $u$, $v$, $w$ and recoil distribution $\rho(x)$ in (\ref{fp}) and (\ref{fq}). Suppose first of all that we have obtained a solution $u_p(x)$ of the equation
\begin{equation}
\label{2ndorder}
-2 \dv[2]{x} u_p(x) = f_p(x) 
\end{equation}
by integrating $f_p(x)$ twice. Then, the corresponding solution to the fourth-order equation (\ref{de}) is
\begin{equation}
\label{pbulk}
p(x) = \sqrt\frac{L}{2\xi} \int_0^1 \dd{x'} u_p(x') {\rm e}^{- \sqrt\frac{2L}{\xi} |x-x'|} \;.
\end{equation}
To see that (\ref{pbulk}) does solve (\ref{de}), we note first that
\begin{equation}
\frac{\xi}{L} \dv[2]{x}\relax {\rm e}^{-\sqrt\frac{2L}{\xi} |x-x'|} =  2 {\rm e}^{-\sqrt\frac{2L}{\xi} |x-x'|} - 2 \sqrt\frac{2\xi}{L} \delta(x-x') \;.
\end{equation}
Then,
\begin{equation}
\frac{\xi}{L} \dv[2]{x} p(x) = 2 p(x) - 2 u_p(x)
\end{equation}
and hence
\begin{equation}
\dv[2]{x} \left( \frac{\xi}{L} \dv[2]{x} -2 \right) p(x)  = -2 \dv[2]{x} u_p(x) = f_p(x)
\end{equation}
as required.

\begin{figure}[tb]
    \begin{center}
        \includegraphics[scale=0.9]{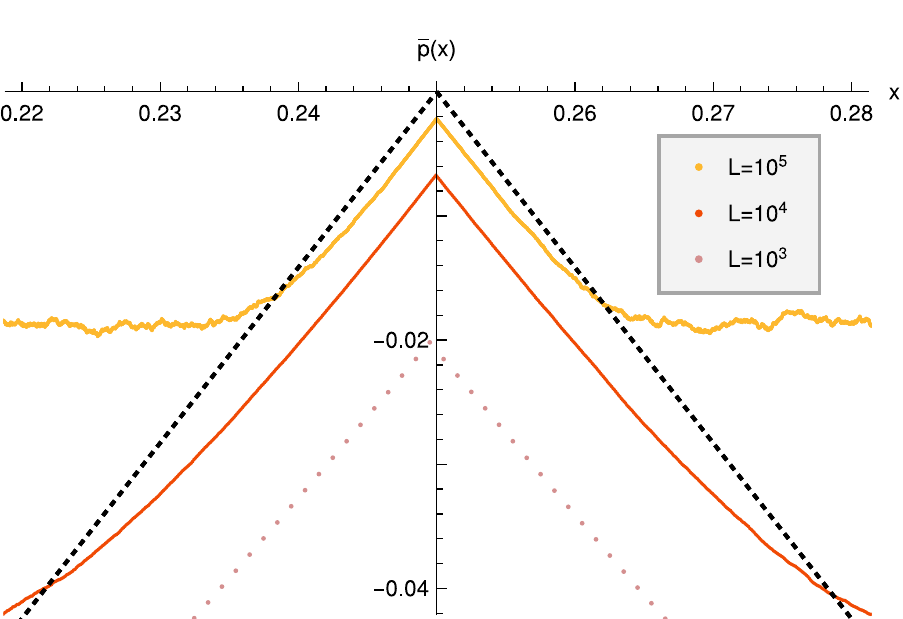}
    \end{center}
    \caption{\label{fig:delta} Limiting behaviour of $\bar{p}(x) = \frac{1}{\sqrt{L}} \ln\left[ \sqrt{\frac{2\xi}{L}} p(x) \right]$ for the recoil distribution $\phi(x) = \delta\left(x-\frac{1}{4}\right)$ and for system parameters $\xi=1$ and $r=1$. The black dashed line corresponds to the $L\to\infty$ limit of $\bar{p}(x)$ which is found from (\ref{deltasm}) as $-\sqrt{\frac{2}{\xi}} \left|x-\frac{1}{4}\right|$. We see the simulation data approaching this limit as $L$ is increased (ordering on plot matches legend ordering).}
\end{figure}

We see then that the effect of the fourth-order term is to convolve the solution $u_p(x)$ of the second-order equation (\ref{2ndorder}) by a function that is sharply peaked over a region of order $\frac{1}{\sqrt{L}}$. When $u_p(x)$ varies smoothly, this `smearing out' of $u_p(x)$ will be barely visible at large $L$. However, if $u_p(x)$ has discontinuities, we will expect to see finite-size corrections within a distance of $\frac{1}{\sqrt{L}}$ of each discontinuity. For example, if $u_p(x)$ contains a delta function at $x=x_0$, i.e., if $u_p(x) \approx u_0 + \Gamma \delta(x-x_0)$ for $x$ close to $x_0$, then in finite-sized systems the stationary distribution will behave as
\begin{equation}
\label{deltasm}
p(x) \approx u_0 + \Gamma \sqrt\frac{L}{2\xi} {\rm e}^{- \sqrt\frac{2L}{\xi} |x-x_0|}
\end{equation}
around $x=x_0$. In Fig.~\ref{fig:delta} we demonstrate this large-$L$ behaviour for a recoil distribution that has a delta function contribution in the bulk. Meanwhile, if $u_p(x)$ has a step from $u_0$ to $u_1$ at $x=x_0$,  it will be smoothed out as
\begin{equation}
p(x) \approx \frac{1}{2} \left[ (u_0+u_1) \pm (u_1-u_0) \left( 1 - {\rm e}^{-\sqrt{\frac{2L}{\xi}} |x-x_0|} \right) \right]
\end{equation}
in which the $+$ sign applies for $x>x_0$ and the $-$ sign for $x<x_0$. 

In principle, we obtain similar finite-size corrections in regions of size $\frac{1}{\sqrt{L}}$ at each boundary. However, it turns out that these do not need to be considered explicitly due to the matching procedure described in Section~\ref{sec:match} below.

It remains to specify the forms of $u_p(x)$ and $u_q(x)$ that are obtained by twice integrating $f_p(x)$ and $f_q(x)$, respectively, from $x=0$. For following particles we find
\begin{widetext}
\begin{multline}
\label{up}
u_p(x) = \frac{r+(1-r)\kappa_L}{2} w\xi \rho_+(x) -\frac{r\kappa_L +(1-r)}{2} w \int_0^x  \rho_-(y) \dd{y} + {}\\
\frac{1+\kappa_L}{2\xi} w \left[(1-x) \int_0^x y \rho_+(y) \dd{y} + x \int_x^1 (1-y) \rho_+(y) \dd{y} \right]
\end{multline}
and for approaching particles that
\begin{multline}
\label{uq}
u_q(x) = - \frac{r\kappa_L +(1-r)}{2}w \int_0^x \rho_-(y) \dd{y} - \frac{1+\kappa_L}{2}w \int_0^x \rho_+(y) \dd{y} + {}\\
\frac{1+\kappa_L}{2\xi} w \left[(1-x) \int_0^x y \rho_{+}(y) \dd{y} + x \int_x^1 (1-y) \rho_+(y) \dd{y}
 \right] \;.
\end{multline}
\end{widetext}

\section{Solution at the boundaries}
\label{sec:bdy}

We now turn our attention to the form of the stationary distribution near the boundaries of the lattice. Of key importance is the parameter $\kappa_L$, which is given by (\ref{eq::prob_ratio}) and involves the generating function of the recoil distribution, $\tilde\Phi(s)$ evaluated at $s=z$. For the recoil distribution (\ref{Phiuvw}) we have
\begin{equation}
\tilde{\Phi}_{\pm}(z) \sim (u\pm v)(z \pm z^{L-1}) + w \int_0^1 \rho_{\pm}(x) z^{Lx} \dd{x} \;.
\end{equation}
For large $L$, the smaller root of (\ref{z}) behaves as
\begin{equation}
z \sim 1 - \sqrt\frac{2}{L\xi} + \order{\textstyle\frac{1}{L}}
\end{equation}
and hence
\begin{equation}
z^{Lx} \sim {\rm e}^{-\sqrt\frac{2L}{\xi} x} \;.
\end{equation}
By noting that the integral is dominated by contributions at the boundaries, we find
\begin{equation}
\tilde{\Phi}_{\pm}(z) \sim (u\pm v)+ \frac{w\xi\rho_{\pm}(0) - 2(u\pm v)}{\sqrt{2L\xi}} + \order{\textstyle\frac{1}{L}} \;.
\end{equation}
When either $r>0$ or $w>0$, we find that (\ref{eq::prob_ratio}) has the large-$L$ expansion
\begin{equation}
\label{kexpand}
\kappa_L \sim \frac{r(1-w)}{r(1-w)+w} +
\frac{\chi_{r,w} + rw \xi \rho_+(0)}{[r(1-w)+w]^2} \frac{1}{\sqrt{2L\xi}} + \order{\textstyle\frac{1}{L}}
\end{equation}
where
\begin{equation}
\label{chi}
\chi_{r,w} = w-r(1-w) - [r(1-w)+w(1-r)](u-v)
\end{equation}
and we have used $u+v=1-w$ to simplify the expressions. This expansion does not apply when $r=w=0$, due to cancellations that occur when $u+v=1$. Then, one instead has the exact result
\begin{equation}
\label{kappa00}
\kappa_L = v + u\frac{1-z}{1+z} \frac{1+z^{L-1}}{1-z^{L-1}}  \sim v + u \frac{1}{\sqrt{2L\xi}} + \order{\textstyle\frac{1}{L}} \;.
\end{equation}
Note particularly that this does not arise as a limit of (\ref{kexpand}), indicating that the case $r=w=0$ is distinct. We thus treat this special case separately (in Section~\ref{sec:special}), focussing in the meantime on the generic case.

Recalling that $\kappa_L$ is defined as the ratio $\frac{P_{++}(1)}{P_{+-}(1)}$, we see that $P_{++}(1)$ vanishes as $L\to\infty$ only for certain special parameter choices. These include the case $w=1$ that was treated in \cite{shortshock}, and where there are no delta function contributions to the recoil distribution at the boundaries. In this case, a number of simplifications occur, including being able to drop the $\kappa_L$ factors that appear in  (\ref{up}) and (\ref{uq}) and the boundary conditions on the discrete distribution $P_{+-}(n)$ carrying over to $q(x)$. It is these simplifications that facilitated the more elementary treatment presented in \cite{shortshock}.

\subsection{Following particles}

To establish the behaviour of the stationary distributions at the boundaries, we return to the original master equations (\ref{me1}) and (\ref{me2}). It is helpful first of all to set the overall normalization of the distribution by putting $P_{+-}(1)=1$. Then, $P_{++}(1) = \kappa_L$, via (\ref{eq::prob_ratio}). From (\ref{me1}) we have for following particles and $n \ll L$ that
\begin{multline}
P_{++}(n+1) - P_{++}(n) = P_{++}(n)-P_{++}(n-1) - {}\\
 (r + (1-r) \kappa_L) (1-w) \delta_{n,1} + \order{\textstyle\frac{1}{L}} \;,
\end{multline}
recalling the boundary condition $P_{++}(0)=0$. For the case $n=1$ (and we do not have $r=w=0$) we obtain
\begin{equation}
P_{++}(2) - P_{++}(1) = \frac{\chi_{r,w} + rw \xi \rho_+(0)}{r(1-w)+w} \frac{1}{\sqrt{2L\xi}} + O({\textstyle\frac{1}{L}}) \;.
\end{equation}
For $n>1$ we have meanwhile
\begin{equation}
P_{++}(n+1) - P_{++}(n) = P_{++}(n)-P_{++}(n-1) + \order{\textstyle\frac{1}{L}} \;.
\end{equation}
Iterating this equation leads us to conclude that near the left boundary,
\begin{equation}
\label{P++bdy}
P_{++}(n) = \kappa_L + \frac{\chi_{r,w} + rw \xi \rho_+(0)}{r(1-w)+w} \frac{n-1}{\sqrt{2L\xi}} 
+ \order{\textstyle\frac{1}{L}} \;.
\end{equation}
That is, on the lattice, the distribution for following particles approaches a value of $\kappa_L$ at the left boundary linearly with a gradient that is proportional to $\frac{1}{\sqrt{L}}$. Due to the symmetry $P_{++}(L-n) = P_{++}(n)$, the behaviour at the right boundary is the same.

\subsection{Approaching particles}

The boundary solution for approaching particles is more complex, as we need to keep terms up to order $\frac{1}{L}$ in this case. The master equation (\ref{me2}) can be written as
\begin{equation}
\label{P+-iter}
P_{+-}(n+1) = (1+\omega) P_{+-}(n) - \lambda(n)
\end{equation}
where
\begin{equation}
\lambda(n) = \omega P_{++}(n) + [r \kappa_L + (1-r)] \Phi(n) \;.
\end{equation}
Iterating from the left boundary, at which $P_{+-}(1)=1$, we find for $n \ll L$ that
\begin{equation}
P_{+-}(n) = (1+\omega)^{n-1} - \sum_{0 < k < n} (1+\omega)^{n-k-1} \lambda(k)  \;.
\end{equation}
Substituting $\omega = \frac{1}{L\xi}$, $\Phi(n)$ from (\ref{Phiuvw}) and $P_{++}(n)$ from (\ref{P++bdy}), and keeping terms up to order $\frac{1}{L}$ in a large-$L$ expansion, we find
\begin{multline}
\label{P+-left}
P_{+-}(n) = 1 - u [r\kappa_L + (1-r)] \left(1 - \frac{1}{L\xi}\right) I_{n>1} + {} \\
 \left( 1-\kappa_L - [r\kappa_L + (1-r)] [u + w \xi \rho(0)]  \right) \frac{n-1}{L\xi}
\end{multline}
in which $I_{n>1}$ is an indicator function, equalling $1$ if $n>1$ and $0$ otherwise. Thus, on the lattice, the stationary distribution for following particles approaches a value close to $1-u[r\kappa_L + (1-r)]$ linearly with a gradient proportional to $\frac{1}{L}$. When $u>0$, there is a step between sites $1$ and $2$. That is, if there is some probability that particles jam on contact, the imposed boundary value $P_{+-}(1)$ is not approached smoothly at the left boundary.

\begin{figure*} 
    \begin{center}
        \includegraphics[width=0.75\linewidth]{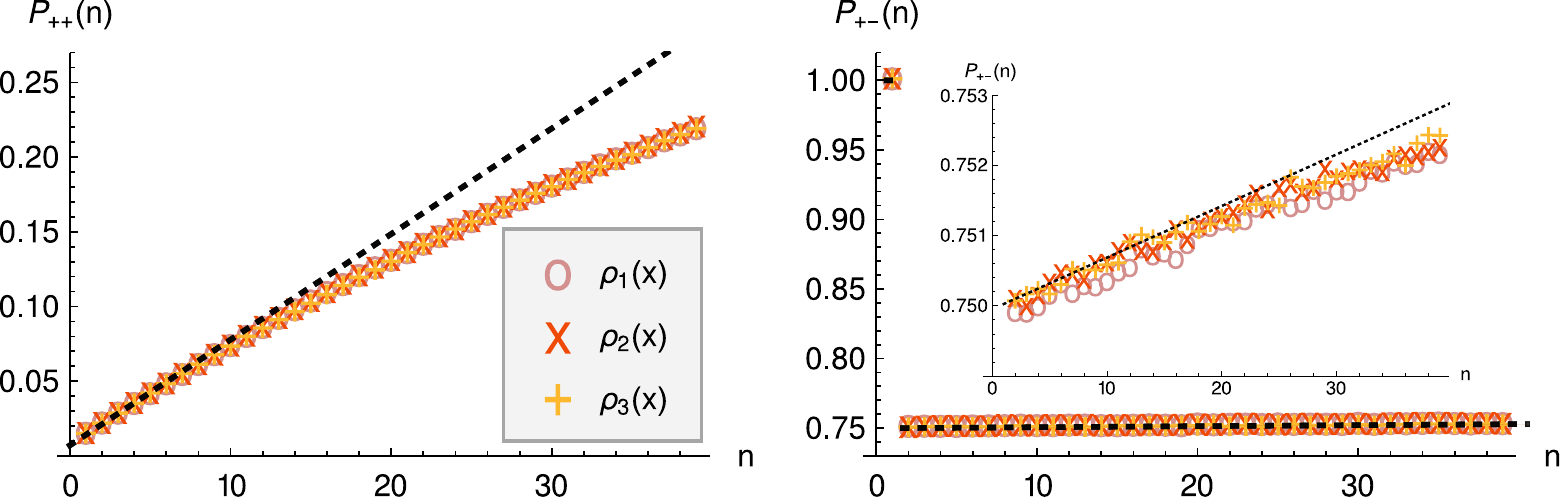}
    \end{center}
    \caption{\label{fig:ubound} Demonstration of the common behavior at the left boundary in the $++$ (left) and $+-$ (right) sectors for three contrasting recoil distributions, with the same boundary terms: $\phi_1(x)=3x(1-x)+\frac{1}{4}\delta(x)+\frac{1}{4}\delta(1-x)$, $\phi_2(x)=\Pi(2x-1)+\frac{1}{4}\delta(x)+\frac{1}{4}\delta(1-x)$, and $\phi_3(x)=\frac{1}{2}\delta(\frac{1}{2}-x)+\frac{1}{4}\delta(x)+\frac{1}{4}\delta(1-x)$. Here, $\Pi(x)$ is defined to be the top-hat function of unit width centred at the origin. System parameters are $L=10^4$, $\xi=1$ and $r=0$. Note that, despite the bulk recoil distributions ranging from smooth to discontinuous, the boundary behaviour is, as predicted, identical in all cases. The slight shift of the pink circles away from the predicted curve in the $+-$ sector is attributed to corrections of higher order than $\frac{1}{L}$ considered in the derivation of (\ref{P+-left}).}
\end{figure*}

We can perform the corresponding analysis at the right boundary by iterating equation (\ref{P+-iter}) in the opposite direction, and using the boundary condition $P_{+-}(L)=0$. In this case we find
\begin{equation}
P_{+-}(L-n) = \sum_{0<k\le n} \frac{\lambda(k)}{(1+\omega)^{n+1-k}}
\end{equation}
when $n\ll L$. Performing the large-$L$ expansion up to order $\frac{1}{L}$ yields
\begin{multline}
\label{P+-right}
P_{+-}(L-n) = v [r\kappa_L + (1-r)] \left(1 - \frac{1}{L\xi}\right) + {} \\
 \left( \kappa_L - [r\kappa_L + (1-r)] [v - w \xi \rho(1)]  \right) \frac{n-1}{L\xi} \;.
\end{multline}
At the right boundary, the distribution approaches a value close to $v [r\kappa_L + (1-r)]$, again linearly and with a gradient proportional to $\frac{1}{L}$.

We note that the solutions near the boundary, (\ref{P++bdy}), (\ref{P+-left}) and (\ref{P+-right}) do not depend on the functional form of the recoil distribution in the bulk. Thus, for any two recoil distributions $\phi_1(x)$ and $\phi_2(x)$, that satisfy $\rho_1(0)=\rho_2(0)$ and $\rho_1(1)=\rho_2(1)$, the corresponding stationary distributions near the boundaries will be the same (as long as the parameters $u$, $v$, $r$ and $\xi$ are also the same). Fig.~\ref{fig:ubound} demonstrates this common boundary behavior across three recoil distributions which are distinct in the bulk but identical at the domain boundaries. Despite the contrasting forms of the bulk distributions, we see a striking universality, as predicted.

\section{Matching of the bulk and boundary solutions}
\label{sec:match}

In Section~\ref{sec:bulk}, we constructed a particular solution of the fourth-order equation (\ref{de}) that applies in the bulk as $L\to\infty$. For following particles, this solution is approximated by $u_p(x)$, given by (\ref{up}), and for approaching particles by $u_q(x)$, given by (\ref{uq}). Since these particular solutions depend on the functional form of the recoil distribution $\rho(x)$ in the bulk, they will not in general match the boundary distributions (\ref{P++bdy}), (\ref{P+-left}) and (\ref{P+-right}), which are universal. To match the solutions, we must add to $u_p(x)$ and $u_q(x)$ solutions of the differential equation (\ref{de}) with a zero right-hand side. These take the form
\begin{equation}
\label{hom}
h(x) = A + B x + C{\rm e}^{-\sqrt\frac{2L}{\xi} x} + D{\rm e}^{-\sqrt\frac{2L}{\xi} (1-x)}
\end{equation}
where the constants $A$, $B$, $C$ and $D$ will be different for following and approaching particles, and need to be chosen such that the correct behaviour is reproduced at the boundaries. Recall that throughout this section we assume that either $r>0$ or $w>0$: the case $r=w=0$ will follow in Section~\ref{sec:special}.

\subsection{Following particles}

We begin with the case of following particles. Since the stationary distribution has the symmetry $p(x)=p(1-x)$, and $u_p(x)$ also exhibits this symmetry, we must have $h_p(x)=h_p(1-x)$. This is achieved when
\begin{equation}
\label{phom}
h_p(x) = A_p + C_p \left(  {\rm e}^{-\sqrt\frac{2L}{\xi} x} + {\rm e}^{-\sqrt\frac{2L}{\xi} (1-x)} \right) \;.
\end{equation}
From (\ref{P++bdy}), we find for $x$ of order $\frac{1}{L}$ that
\begin{equation}
p(x) \sim \kappa_L + \frac{\chi_{r,w} + rw \xi \rho_+(0)}{r(1-w)+w}  \sqrt\frac{L}{2\xi} \left(x - \frac{1}{L}\right) \;.
\end{equation}
Adding (\ref{phom}) to (\ref{up}) we also have
\begin{equation}
\label{pbdy}
p(x) \sim \frac{r+(1-r)\kappa_L}{2} w\xi\rho_{+}(0) + A_p + C_p \left(1 - \sqrt\frac{2L}{\xi}x\right)
\end{equation}
in this boundary region. Comparing coefficients of $x$ in these two expressions implies that
\begin{equation}
C_p = - \frac{1}{2} \frac{\chi_{r,w} + rw \xi \rho_+(0)}{r(1-w)+w} \;.
\end{equation}
Comparing the constant terms, and using (\ref{kexpand}), we find that
\begin{equation}
A_p = \frac{1}{2} \frac{2r(1-w)+\chi_{r,w}}{r(1-w)+w} + \order{\textstyle\frac{1}{\sqrt{L}}} \;.
\end{equation}
If we want the bulk solution to match the boundary form (\ref{P++bdy}) exactly, we should include the term of order $\frac{1}{\sqrt{L}}$ in $A_p$. Our main interest here is to identify what happens as $L\to\infty$, for which the leading terms identified above are sufficient.

From (\ref{pbdy}) we see that for $x \gg \frac{1}{\sqrt{L}}$ the bulk solution tends towards
\begin{equation}
\frac{1}{2} \frac{rw\xi\rho_+(0) + 2r(1-w) + \chi_{r,w}}{r(1-w)+w}
\end{equation}
as either boundary is approached. Within the boundary regions, where $x$ or $1-x$ is of order $\frac{1}{\sqrt{L}}$, the stationary distribution varies exponentially with $x$, reaching $\frac{r(1-w)}{r(1-w)+w}$ at $x=0$ and $x=1$. We see that, generically, the values of $p(x)$ at the boundary point $x=0$ and $x\sim \frac{1}{\sqrt{L}}$ are different, and therefore the exponential feature at the left boundary sharpens to a step as $L\to\infty$. The same behavior is seen at the right boundary (necessarily, due to the symmetry). In summary, the inter-particle distribution function $p(x)$ for following particles is discontinuous at the boundary points $x=0$ and $x=1$. 

In the special case $w=1$ that was treated in \cite{shortshock}, we find that $p(0)=p(1)=0$, consistent with the boundary conditions (\ref{mebc1}) on the original master equation. When the recoil distribution has delta function contributions at the boundaries $p(x)$ assumes a nonzero value at $x=0$ and $x=1$, indicating that one cannot in general directly apply the boundary conditions on the discrete equations to their continuum counterparts.

\subsection{Approaching particles}

For the case of approaching particles, the distribution $q(x)$ does not have any particular symmetry, and we require all four terms in (\ref{hom}). At the left boundary, we have from (\ref{P+-left})
\begin{multline}
\label{qleft}
q(x) \sim 1 - u [r\kappa_L + (1-r)] + {} \\
 \left( 1-\kappa_L - [r\kappa_L + (1-r)] [u + w \xi \rho(0)]  \right) \frac{x}{\xi}
\end{multline}
for $x$ of order $\frac{1}{L}$, but sufficiently large that we avoid the step between the first and second lattice sites. In principle we should retain all terms up to order $\frac{1}{L}$ if we want the bulk and boundary solutions to exactly match, but again we can dispense with these if our aim is to understand the general nature of the stationary distribution in the limit $L\to\infty$. By adding (\ref{hom}) to (\ref{uq}) we find, within the same boundary regime and level of approximation, that
\begin{equation}
\label{uhleft}
q(x) \sim u_q'(0) x + A_q + B_q x + C_q \left( 1 - \sqrt\frac{2L}{\xi} x\right)
\end{equation}
since $u_q(0) = 0$.  At the right boundary, we have from (\ref{P+-right}) that
\begin{multline}
\label{qright}
q(1-x) \sim  v [r\kappa_L + (1-r)] + {} \\
 \left( \kappa_L - [r\kappa_L + (1-r)] [v - w \xi \rho(1)]  \right) \frac{x}{\xi}
\end{multline} 
and by adding (\ref{hom}) to (\ref{uq}) that
\begin{multline}
\label{uhright}
q(1-x) \sim -(1+\kappa_L) w  - u_q'(1) x + {}\\
A_q + B_q(1-x) +  D_q  \left( 1 - \sqrt\frac{2L}{\xi} x\right) 
\end{multline}
when $x$ is of order $\frac{1}{L}$. We recall that $\kappa_L$ is order $1$ when $w\ne1$ and of order $\frac{1}{\sqrt{L}}$ when $w=1$.

Since there are no contributions of order $\sqrt{L}$ in (\ref{qleft}) or (\ref{qright}), it follows that $C_q$ and $D_q$ must at most of order $\frac{1}{\sqrt{L}}$. Comparing the constant terms in (\ref{qleft}) and (\ref{uhleft}), we find to leading order that
\begin{equation}
A_q = 1 - u [r\kappa_L + (1-r)] \;.
\end{equation}
A similar comparison of (\ref{qright}) and (\ref{uhright}) implies that, to the same order, $B_q=0$. Then, by comparing the coefficients of $x$ in (\ref{qleft}) and (\ref{uhleft}), and performing some algebra, we find
\begin{equation}
C_q = - \frac{1}{2} \frac{\chi_{r,w}+ r w \xi \rho_+(0)}{r(1-w)+w} \frac{1}{\sqrt{2L\xi}}
\end{equation}
in which $\chi_{r,w}$ is given by (\ref{chi}). The same procedure applied to (\ref{qright}) and (\ref{uhright}) yields
\begin{equation}
D_q = \frac{1}{2} \frac{\chi_{r,w}+ r w \xi \rho_+(0)}{r(1-w)+w} \frac{1}{\sqrt{2L\xi}} \;.
\end{equation}

In this sector, the behavior at the boundaries is rather complex. As previously noted, the solution (\ref{P+-left}) near the left boundary on the lattice steps from a value of $1$ at $n=1$ to a value close to $1-u[r\kappa_L+(1-r)]$ which is different from $1$ when $u>0$. This step carries through to the continuum solution: we have $q(0)=1$ but
\begin{equation}
\lim_{x\to0+} q(x) = 1 - u \frac{r(1-w) + (1-r)w}{r(1-w) + w} \;.
\end{equation}
Similarly, at the right boundary $q(1)=0$ but from (\ref{P+-right}) we have
\begin{equation}
\lim_{x\to1-} q(x) = v \frac{r(1-w) + (1-r)w}{r(1-w) + w}
\end{equation}
which vanishes only if $v=0$. Thus, except in the case $w=1$ in which there are no delta function contributions to the recoil distribution, the inter-particle distribution function for following particles has steps at both boundaries. These steps are sharp even at finite system sizes $L$, as indicated by the discrete solutions (\ref{P+-left}) and (\ref{P+-right}) not smoothly approaching the boundary conditions (\ref{mebc2}) on the master equation. Again, these boundary conditions on the discrete distribution do not carry over to the continuum.

\begin{figure}[tb]
    \begin{center}
        \includegraphics[scale=0.9]{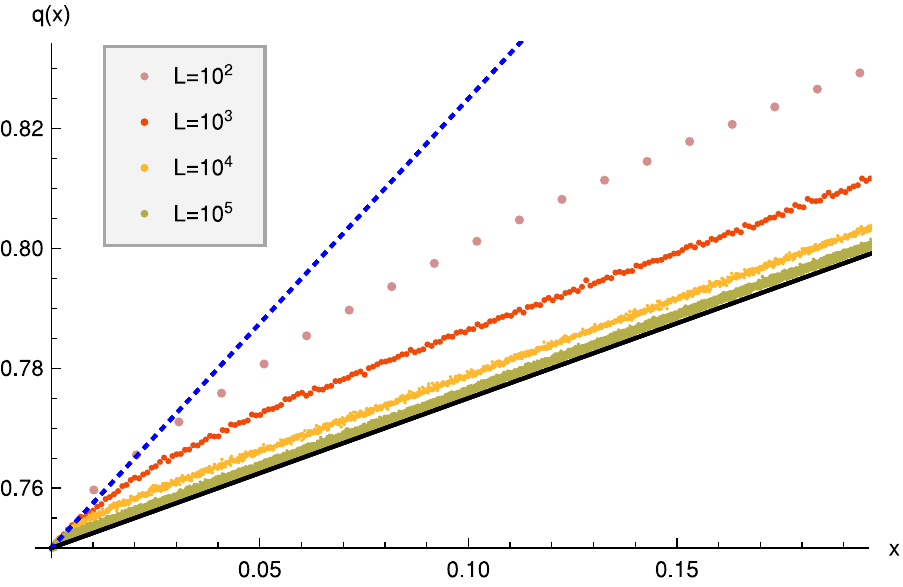}
    \end{center}
    \label{fig:discderiv}
    \caption{Behavior near the left boundary in the $+-$ sector. Simulation data were obtained for different $L$ (ordering on plot matches legend ordering) with the recoil distribution $\phi(x)=\Pi(2x-1)+\frac{1}{4}\delta(x)+\frac{1}{4}\delta(1-x)$ and for system parameters $\xi=1$ and $r=0$. $\Pi(x)$ is the top-hat function of unit width centred at the origin. As $L\to\infty$ the derivative of $q(x)$ should, at any fixed $x$, approach that of the solid black line which is given by (\ref{bulkderiv}). At finite $L$, the derivative of $q(x)$ should approach that of the dashed blue line which is given by the distinct expression, (\ref{boundaryderiv}). Note that for $L=100$ there are higher-order corrections in $L$ which obscure the latter limit.}
\end{figure}

The derivative of $q(x)$ is discontinuous at the boundaries, even when $w=1$. To see this, we consider first the value of the derivative that the solution in the bulk approaches as the coordinate $x\to0$. This is obtained by first taking the limit $L\to\infty$ (causing the size of the boundary region to shrink) at fixed $x$, and then taking $x\to0$ from above. From this process, we find
\begin{multline}
  \label{bulkderiv}
q'(0) = \frac{w}{2\xi} \frac{1}{r(1-w)+w} \Big( 2r(1-w) + w - {}\\ 
  2[r(1-w)+w(1-r)]\xi\rho(0) - r\xi\rho_+(0) \Big) \;.
\end{multline}
If we reverse the order of limits, the limit $x\to0^+$ enters the boundary layer first, which is then subsequently shrunk. This gives us the derivative that applies to the solution at the boundary, rather than the one in the bulk, and we find
\begin{equation}
\label{boundaryderiv}
q'(x) = \frac{1}{\xi} \frac{w - [r(1-w)+w(1-r)] [u + w \xi \rho(0)]}{r(1-w)+w}  \;.
\end{equation}
It is straightforward to see that these two expressions are not, in general, equal. For example, (\ref{bulkderiv}) depends on $\rho(1)$ via $\rho_+(0)=\rho(0)+\rho(1)$, whilst (\ref{boundaryderiv}) does not. Meanwhile, (\ref{boundaryderiv}) depends on the probabilities $u$ and $w$ separately, whilst (\ref{bulkderiv}) depends only on $w$. This behavior is confirmed by Fig.~\ref{fig:discderiv}, where we find (\ref{bulkderiv}) by looking at $L\to\infty$ at finite $x$, and (\ref{boundaryderiv}) by looking at $x\to0$ at finite $L$. One finds a similar behavior at the right boundary.

These results indicate that one cannot apply boundary conditions that derive from the correct behavior in the boundary layer directly to the continuum limit of the bulk solution. Instead, one has to retain finite-size corrections to the bulk solution, and apply the boundary conditions before taking the continuum limit.

\section{Partially jamming particles with no recoil-induced reversals}
\label{sec:special}

We finally turn to the special case of $r=w=0$, where the expansion (\ref{kexpand}) for $\kappa_L$ does not apply, and we need to use the exact form (\ref{kappa00}) instead. This case corresponds to a dynamics in which, whenever a particle attempts to hop on top of the other, they return to their original positions with probability $u$ (`jam') or exchange places with probability $v=1-u$. In both cases, the velocity configuration is unchanged by recoil: there are no recoil-induced reversals. We can view these as particles which partially jam on contact, and otherwise can move through one another.

It turns out that this case can be solved exactly on the lattice without too much difficulty. The key is to note that in the bulk, the right-hand side of the fourth-order difference equation (\ref{discrete4}) vanishes, and the homogeneous solution of that equation, which reads
\begin{equation}
\label{Hn}
H(n) = A + B \frac{n}{L} + C z^n + D z^{L-n}
\end{equation}
applies to both $P_{++}(n)$ and $P_{+-}(n)$ sufficiently far from the boundaries. From the analysis of Section~\ref{sec:bdy} we know that there is a step between sites $n=1$ and $n=2$ in the distribution for approaching particles. For following particles we also have the symmetry $P_{++}(n)=P_{++}(L-n)$. These observations lead to the ans\"{a}tze
\begin{align}
P_{++}(n) &= A + C(z^n + z^{L-n}) \\
P_{+-}(n) &= A' + B'n + C' z^n + D' z^{L-n} + \Delta \delta_{n,1} \;.
\end{align}

Substituting these expressions into the master equation (\ref{me2}), we find that they can only hold for all $n$ satisfying $\Phi(n)=0$ if
\begin{equation}
A-A' = B' = 0 \qand C' = - D' = \frac{1-z}{1+z} C \;.
\end{equation}
This leaves us to determine $A$, $C$ and $\Delta$, which can be achieved by solving the linear system formed by the three equations $P_{+-}(1)=1$, $P_{++}(1)=\kappa_L$ and (\ref{P+-iter}) at $n=L-1$, recalling that we have the boundary condition $P_{+-}(L)=0$. 

Using the exact expression (\ref{kappa00}) for $\kappa_L$ we can solve the linear system for $A$, $B$ and $C$, finding
\begin{align}
A &= v + u \frac{(1-z)^2}{(1+z)(1+z^2)} \frac{1-z^L}{1-z^{L-1}} \\
C &= u \frac{1-z}{(1+z^2)(1-z^{L-1})} \\
\Delta &= \frac{2uz}{1+z^2} \;.
\end{align}
These expressions agree with the exact result for $u=1$ given in \cite{Slowman2016}.

We may now take the limit $L\to\infty$ with $z \sim 1 - \sqrt{\frac{2}{L\xi}}$ to obtain
\begin{equation}
\label{pspecial}
p(x) = v + \frac{u}{2L\xi} + \frac{u}{\sqrt{2L\xi}} \left( {\rm e}^{-\sqrt{\frac{2L}{\xi}} x} + {\rm e}^{-\sqrt{\frac{2L}{\xi}} (1-x)} \right) \;.
\end{equation}
The terms that ultimately contribute depend on the value of the probability $v$ that particles pass through each other on contact. If $v$ retains some fixed non-zero value in the limit $L\to\infty$, the inter-particle distribution function $p(x)$ is constant across the entire interval $x\in[0,1]$. This result can be understood fairly straightforwardly: if the particles initially jam (with probability $u=1-v$), it is likely that one will attempt to hop on top of the other before a velocity reversal occurs, and eventually one of these hop attempts will cause the particles to exchange places. Thus the two particles are essentially invisible to each other, which leads to the uniform distribution over their separation.

When $v=0$ (and thus $u=1$), all terms in $u$ in (\ref{pspecial}) survive, despite being superficially of different orders in $L$. This is because, for $x\in[0,1]$, the combination $\sqrt{L} {\rm e}^{-\sqrt{L} x} \to \delta(x)$ as $L\to\infty$. For the case $v=0$, we find
\begin{equation}
\label{pv0}
p(x) = \frac{1}{\xi} + \left[ \delta(x) + \delta(1-x) \right] \;,
\end{equation}
in agreement with the result of \cite{Slowman2016}. We can further identify a crossover regime, where the jamming probability scales as $\frac{1}{L}$ and all three terms in (\ref{pspecial}) contribute. Specifically, taking $v = \frac{\hat{v}}{2L}$, we obtain
\begin{equation}
\label{pv}
p(x) = \hat{v} + \frac{1}{\xi} + \left[ \delta(x) + \delta(1-x) \right]  \;.
\end{equation}
This tells us that the accumulation of probability at the boundary points $x=0$ and $x=1$ (indicated by the delta functions) is not a generic feature of the inter-particle distribution, but appears only when the probability $v$ that the particles exchange places on contact is of order $\frac{1}{L}$ or smaller. In this regime there is some nonzero probability that a reversal takes place before the particles pass through each other.

For the case of approaching particles, we see in the large-$L$ limit that
\begin{multline}
\label{qspecial}
q(x) = v + \frac{u}{2L\xi} + \frac{u}{2L\xi} \left( {\rm e}^{-\sqrt{\frac{2L}{\xi}} x} - {\rm e}^{-\sqrt{\frac{2L}{\xi}} (1-x)} \right) \\
{} + u \Theta({\textstyle\frac{1}{L}}-x)
\end{multline}
where $\Theta(\cdot)$ is the step function that is zero for negative argument and unity otherwise. Again, for any fixed, nonzero $v$ and $x>0$ we find that $q(x)=v$ as $L\to\infty$.  With $v=\frac{\hat{v}}{2L}$ we find
\begin{equation}
\label{qv}
q(x) = \hat{v} + \frac{1}{\xi} + 2\delta(x) \;.
\end{equation}
Again this agrees with \cite{Slowman2016} for the case $\hat{v}=0$.

\section{Summary of results}
\label{sec:summary}

In this section, we summarize the main results of this work. Since the focus of this work has been the continuum limit, we neglect here to show results for the discrete case, where we instead refer the reader to section \ref{sec:sss} and appendix \ref{sec:urgh}.

We considered recoil distributions of the form
\begin{equation}
    \phi(x) = u\delta(x) + v\delta(1-x) + w\rho(x) \;,
\end{equation}
where $u+v+w=1$. In this expression, the first contribution corresponds to jamming, the second contribution corresponds to particle exchange, and the third contribution is a normalised distribution on the interval $0\leq x\leq 1$ which is assumed continuous at the boundaries but is permitted to be discontinuous elsewhere. We showed that the two stationary distributions $p(x)$ and $q(x)$ for following and approaching particles, respectively, obey the fourth-order differential equation
\begin{equation}
    \dv[2]{x} \left( \frac{\xi}{L} \dv[2]{x} -2 \right) \left(\begin{array}{c} p(x) \\ q(x)  \end{array}\right) = \left(\begin{array}{c} f_p(x) \\ f_q(x)  \end{array}\right) \;,
\end{equation}
where $f_p(x)$ and $f_q(x)$ are source terms given in (\ref{fp}) and (\ref{fq}). The solution to this equation was found by adding complementary terms to a particular solution and subsequently matching their boundary behaviours to limiting forms of the discrete solutions near the domain boundaries. This led to the results
\begin{widetext}
\begin{multline}
\left(
\begin{array}{c}
    p(x) \\
    q(x)
\end{array}
\right)
=
\left(
\begin{array}{c}
    A_p \\
    A_q
\end{array}
\right)
+
\left(
\begin{array}{c}
    B_p \\
    B_q
\end{array}
\right)
x
+
\left(
\begin{array}{c}
    C_p \\
    C_q
\end{array}
\right)
{\rm e}^{-\sqrt{\frac{2L}{\xi}} x}
+
\left(
\begin{array}{c}
    D_p \\
    D_q
\end{array}
\right)
{\rm e}^{-\sqrt{\frac{2L}{\xi}} (1-x)}
+
\sqrt\frac{L}{2\xi} \int_0^1 \dd{x'}
\left(
\begin{array}{c}
    u_p(x') \\
    u_q(x')
\end{array}
\right)
{\rm e}^{- \sqrt\frac{2L}{\xi} |x-x'|} \;.
\end{multline}
\end{widetext}
where $u_p(x)$ and $u_q(x)$ are as stated in (\ref{up}) and (\ref{uq}), and where the constants of integration are
\begin{align}
    A_p &= \frac{1}{2} \frac{2r(1-w)+\chi_{r,w}}{r(1-w)+w} \\
    C_p &= D_p = -\frac{1}{2} \frac{\chi_{r,w} + rw \xi \rho_+(0)}{r(1-w)+w} \\
    A_q &= 1 - u [r\kappa_L + (1-r)] \;. 
\end{align}
Here $\chi_{r,w}$ depends on the model parameters as
\begin{equation}
    \chi_{r,w} = w-r(1-w) - [r(1-w)+w(1-r)](u-v) .
\end{equation}
All other constants of integration vanish at $\order{1}$.

Due to leading-order cancellations, the case of partially-jamming particles with no recoil-induced velocity reversals ($r=w=0$) is easiest to treat by first finding the discrete solution and subsequently taking the continuum limit. The results here, it turns out, are very sensitive to the particle-exchange probability $v$. Most interestingly, when $v=\frac{\hat{v}}{2L}$ for $\hat{v}=\mathcal{O}(1)$, a varied structure comprising uniform and jamming contributions emerges:
\begin{align}
    p(x) &= \hat{v} + \frac{1}{\xi} + \left[ \delta(x) + \delta(1-x) \right]  \; \\
    q(x) &= \hat{v} + \frac{1}{\xi} + 2\delta(x) \;,
\end{align}

The remaining distributions are stated for different orders of $v$ in section \ref{sec:special}.

\section{Discussion}
\label{sec:disco}

In this work we have considered a model system comprising two persistent particles that may jam, pass through each other or recoil on contact. Our starting point was a lattice-based stochastic process, for which the master equations comprise a system of coupled first- and second-order difference equations, (\ref{me1}) and (\ref{me2}). By applying the kernel method, we transformed these to a set of decoupled fourth-order difference equations (\ref{discrete4}) whose continuum limit is a fourth-order differential equation. The solution of these equations involves integrals over the recoil distribution, (\ref{up}) and (\ref{uq}), which arise from  advective and diffusive processes that were shown in \cite{shortshock} to combine nontrivially in generating effective attractive or repulsive interactions.

A feature of the decoupled differential equation is that the coefficient of the fourth-order term is of order $\frac{1}{L}$ relative to the second-order term. Although one might expect to be able to neglect this higher order term in general, the leading finite-size corrections need to be retained to apply the boundary conditions on the general solution of the differential equation.

From the perspective of constructing many-body descriptions of interacting persistent particles, our most significant finding is the highly nontrivial nature of the boundary conditions. For example, the original master equations suggest a vanishing boundary condition on the distribution for following particles. By solving the discrete model in the boundary region, we find from (\ref{P++bdy}) and (\ref{kexpand}) that this is only the case when particles always recoil on contact ($w=1$), or never reverse direction after recoiling ($r=0$). Even when the distribution that emerges in the continuum limit does vanish at the boundaries, it does so over a region of size $\frac{1}{\sqrt{L}}$ implying that there is a step-function discontinuity at the boundaries. Thus we find that it is never appropriate to apply the na\"ive condition that the distribution function vanishes at the boundaries, and doing so would yield an incorrect result.

Similarly, the inter-particle distribution for approaching particles does not smoothly approach the na\"ive values of 1 at the left boundary (arising from normalization) and 0 at the right boundary (as is the case for the discrete master equation) when there is a delta function contribution to the recoil distribution at the relevant boundary. Even when these delta functions are absent, the derivative of this distribution is discontinuous at each boundary. This is significant because if one were able to obtain a boundary condition by applying to physical principles (for example, a zero-flux condition), one would not know whether it should apply to the left or the right of the discontinuity. This in turn could generate an incorrect result. Therefore, if general physical principles do exist for deriving boundary conditions on continuum equations for persistent interacting particles, it seems likely that they will be subtle. The danger of making ad-hoc assumptions on the  boundary or initial conditions is further illustrated by an analysis of a version of the telegrapher equations that generates unphysical solutions when this is done \cite{Tilles2019}. 

We did find, however, that when the recoil distribution is differentiable at the boundary points, the na\"{\i}ve boundary conditions on the inter-particle distribution for approaching particles do carry over from the discrete master equation to the continuum limit. Then, the ambiguity around the application of the boundary conditions disappears, and one obtains the correct answer by neglecting subleading derivatives in the following sector. This was the procedure followed in \cite{shortshock} and the more careful analysis presented here justifies the assumptions that were made in that specific case. However, such a justification was only possible in retrospect, having knowledge of the full solution.

We further found that the limit $r=w=0$, where particles may only jam or pass through each other, is singular. This was initially evident from the expansion (\ref{kexpand}) breaking down in this limit, and the corresponding expansion (\ref{kappa00}) taking a fundamentally different form. In the continuum limit, we find the inter-particle distribution functions are uniform unless the probability $v$ that particles pass through each other vanishes at least as fast as $\frac{1}{L}$ as $L\to\infty$. It is not obvious why this case is distinct, and hints that it may have special mathematical properties (such as some form of integrability, perhaps).

This study has been restricted to the case of a pair of particles. Ideally, we would like to be able to use the knowledge gleaned from the two-body problem to construct a faithful description of an arbitrary number of active particles. This remains an outstanding challenge. In particular, it has been shown that a two-body effective potential is insufficient to generate motility-induced phase separation in active Brownian particles and that many-body terms are required \cite{Turci2021}. Further insights likely require a solution of the master equations (\ref{me1}) and (\ref{me2}) generalized to at least three particles, which has proved challenging. It is possible that simplifications occur when thermal noise is added to the particles' persistent motion. A hint that this might be the case is that the full dynamical spectrum obtained in the discrete formulation for a pair of hard-core persistent particles \cite{Mallmin2019} was subsequently reproduced in the zero-temperature limit of the process with additional thermal noise \cite{Das2020}. Intriguingly, the latter results were found by applying more natural zero flux conditions at the boundaries. We also note that field-theoretic methods have been usefully applied to individual particles undergoing combined persistence and thermal diffusion in an external potential \cite{Garcia2021}, which may generalize more naturally to the many-body case. Such additional insights are likely an essential component of a bottom-up theory for active matter that is fully grounded in microscopic interactions, but at present further work of the kind mentioned above is needed to establish this.

\section*{Acknowledgments}

MJM acknowledges studentship funding from EPSRC through the Scottish CM-CDT under Grant No.~EP/L015110/1. For the purpose of open access, the author has applied a Creative Commons Attribution (CC BY) licence to any Author Accepted Manuscript version arising from this submission.

\appendix

\section{Direct inversion of the generating function}
\label{sec:urgh}

As noted in the main text, it is possible to invert the generating functions $G_{\sigma_1\sigma_2}(s)$ that appear in Eq.~(\ref{linsol}). This is achieved by first rewriting (\ref{linsol}) as
\begin{equation}
    \left(\begin{array}{c} G_{++}(s) \\ G_{+-}(s) \\ G_{-+}(s) \end{array}\right) = -\frac{1}{1+\omega} \frac{s^2\mat{A}(s)\vec{b}(s)}{(1-z^{-1}s)(1-zs)(1-s)^2} \;.
\end{equation}
To write the right-hand side as a power series in $s$, we perform the expansion
\begin{equation}
    \frac{s^2}{(1-z^{-1}s)(1-zs)(1-s)^2} = \sum_{n=0}^\infty c(n) s^n
\end{equation}
to obtain
\begin{equation}
    c(n) = \frac{z^2}{(z^2-1)(z-1)^2} \bigg[ \left( z^{n} - \frac{1}{z^{n}} \right) - 
    n\left( z - \frac{1}{z} \right) \bigg] \;.
\end{equation}
Note that $c(0)=c(\pm1)=0$.
We now take the Cauchy product of the above power series with the polynomials given by the elements of $\mat{A}(s)\vec{b}(s)$, after which the coefficients in $G_{++}(s)$ are read off to reveal
\begin{multline}
    P_{++}(n) = \kappa_L \Delta^2 c(n) - {\cal A} \sum_{m=0}^{n} c(n-m) \Delta^2 \Phi_+(m)  \\
	{} - \frac{\omega}{1+\omega} \left[ \Delta_- c(n) + \omega (1+\kappa_L) c(n) \right] \\
    {} + \frac{\omega^2}{1+\omega} (1+\kappa_L) \sum_{m=0}^{n} c(n-m) \Phi_+(m) \\
    {} + \frac{\omega}{1+\omega} {\cal B} \sum_{m=0}^n c(n-m) \left[ \Delta_- \Phi(m) - \Delta_+ \Phi^*(m) \right]
    \label{P++(n)}
\end{multline}
where
\begin{align}
	\Phi^*(n) &= \Phi(L-n) \\
	\Phi_{+}(n) &= \Phi(n) + \Phi^*(n) \\
    {\cal A} &= r + (1-r)\kappa_L \\
    {\cal B} &= (1-r) + r \kappa_L
\end{align}
and the finite difference operators $\Delta_{\pm}$ and $\Delta^2$ have the same meaning as in Section~\ref{sec:ctm}.

Meanwhile, the coefficients of $G_{+-}(s)$ yield
\begin{multline}
    P_{+-}(n) = \Delta_- \Delta^2 c(n)- {\cal B} \sum_{m=0}^{n} c(n-m) \Delta_- \Delta^2 \Phi(m) \\
    {} - \frac{\omega}{1+\omega} \left[ \Delta_- c(n-1) + (1+\kappa_L) (\omega + \Delta_-) c(n) \right] \\
    {} + \frac{\omega^2}{1+\omega} (1+\kappa_L) \sum_{m=0}^{n} c(n-m) \Phi_+(m) \\
    {} + \frac{\omega}{1+\omega} {\cal A} \sum_{m=0}^{n} c(n-m)  \Delta_- \Phi_+(m) \\
    {} + \frac{\omega}{1+\omega} {\cal B} \sum_{m=0}^{n} c(n-m) \Delta_- \left[\Phi(m) + \Phi(m-1)\right]
    \label{P+-(n)}
\end{multline}

Since the sums in (\ref{P++(n)}) and (\ref{P+-(n)}) are finite, one can readily evaluate them numerically, which was the procedure used to generate the data for Fig.~\ref{fig::test_distribution_discrete} in the main text.  They are, however, much harder to work with than the expressions obtained in the main text directly within the scaling limit.

\end{document}